\documentclass[%
 reprint,
superscriptaddress,
 amsmath,amssymb,
]{revtex4-2}

\usepackage{graphicx}
\usepackage{dcolumn}
\usepackage{bm}
\usepackage{soul}

\usepackage[counterclockwise, figuresleft]{rotating}
\usepackage{multirow}
\usepackage{longtable}
\usepackage{enumitem}

\begin{document}
\onecolumngrid
\title{A weighted multidetector extension to the 5-vector method for the search of continuous gravitational wave signals}

\author{Luca D'Onofrio} \email{ldonofrio@roma1.infn.it}
\affiliation{INFN, Sezione di Roma, I-00185 Roma, Italy} 
\author{Rosario De Rosa}
\affiliation{Università di Napoli “Federico II”, I-80126 Napoli, 
Italy}\affiliation{INFN, Sezione di Napoli, I-80126 Napoli, Italy} 
\author{Cristiano Palomba}
\affiliation{INFN, Sezione di Roma, I-00185 Roma, Italy}


\begin{abstract}
The 5-vector method is one of the main pipeline used by the LIGO-Virgo-KAGRA Collaboration for the analysis of continuous gravitational waves (CWs). This method is a matched filter in the frequency domain based on the splitting in five frequencies of the expected gravitational wave frequency due to the Earth sidereal motion. The data 5-vector is composed by five components, each defined as the Fourier transform of the data at one of the five frequencies where the signal power is split. When dealing with a scenario involving $n$ detectors, the standard multidetector extension combines the data 5-vector from each detector, resulting in a total of $5n$ components. In this letter, we introduce a novel multidetector extension that goes beyond the standard approach. It takes into consideration the varying noise levels and observation times associated with each detector, offering a more sophisticated and effective approach for CWs analysis.
\end{abstract}
\maketitle
\onecolumngrid
\section{Introduction}
The 5-vector method \cite{2010} is a frequentist pipeline used by the LIGO-Virgo-KAGRA Collaboration for the analysis of continuous gravitational waves (CW) that could be emitted by spinning neutron stars with a non-axisymmetric mass distribution. In the source frame, CW signal are quasi-monochromatic with the gravitational wave frequency $f_{gw}$ that is proportional to the source rotation frequency according to the emission model \cite{ornella}. 

In the 5-vector formalism, the CW signal $h(t)$ at the detector can be written as \cite{2010}:
\begin{equation}\label{compsign}
h(t)=H_0(H_+ A_+ + H_\times A_\times)e^{i\Phi(t)}
\end{equation}where\begin{equation}
H_+=\frac{\cos(2\psi)-j\eta \sin(2\psi)}{\sqrt{1+\eta^2}} \qquad \text{and} \qquad H_\times=\frac{\sin(2\psi)+j\eta \cos(2\psi)}{\sqrt{1+\eta^2}} \,.
\end{equation}
$\eta$ and $\psi$ are the polarization parameters and $H_0$ the amplitude. The two functions $A_{+/\times}$ entail the detector response to the coming CW signal \cite{2010}. 

The phase $\Phi(t)$ in \eqref{compsign} shows a time dependence due to different phenomena (spin-down, Doppler effects) that modulate in time the received signal frequency. After the Doppler and spin-down corrections (see \cite{2019}), there is a residual modulation in amplitude and phase due to the Earth sidereal modulation that splits the signal into the 5 frequencies $f_{\text{gw}},f_{\text{gw}} \pm \Omega_\oplus,f_{\text{gw}} \pm 2 \Omega_\oplus $ where $\Omega_\oplus$ is the Earth's sidereal angular frequency.  

The data 5-vector $\textbf{X}$ and the signal template 5-vectors  $\textbf{A}^{+/\times}$ are defined as the Fourier transforms of the data and of the template functions $A_{+/\times}$ at the 5 frequencies where the signal power is split. 
The 5-vector method defines two matched filters between the data $\textbf{X}$ and the signal templates $\textbf{A}^{+/\times}$ vectors, used in order to maximize the signal-to-noise ratio:
\begin{equation}\label{estim}
\hat{H}_+=\frac{\textbf{X}\cdot \textbf{A}^+}{|\textbf{A}^+|^2} \qquad \text{and} \qquad \hat{H}_\times=\frac{\textbf{X}\cdot \textbf{A}^\times}{|\textbf{A}^\times|^2}
\end{equation} These two matched filters are used \cite{2010} to estimate the signal parameters and to construct the detection statistic. In this letter, we first review the standard multidetector extension of the 5-vector and then, propose a new method that considers the different detectors' noise levels and observation times.

\section{Standard multidetector extension}\label{subsec:5n}
Let us consider a network of $n$ detectors and for the j-th detector, let us compute the corresponding signal $\mathbf{X_j}$ and template 5-vectors $\mathbf{ A^{+/\times}_j}$. In \cite{5n}, the 5n-vectors are defined as:
\begin{equation}\label{5nvec}
\mathbf{X}=[\mathbf{ X_1},..., \mathbf{ X_n}] \,,
\qquad
\mathbf{ A^+}=[\mathbf{ A^+_1},..., \mathbf{ A^+_n}] \,,
\qquad
\mathbf{ A^\times}=[\mathbf{ A^\times_1},..., \mathbf{ A^\times_n}]
\end{equation}combining the data 5-vectors $\mathbf{ X_j}$ and the template 5-vectors $\mathbf{ A^{+/x}_j}$ (with $j=1,.., n$) for the considered pulsar in the j-th detector. 
The multi-detector statistic $S$ (as defined in \cite{mio2}) is:
\begin{equation}\label{statS}\small
S= \frac{|\textbf{A}^{+}|^4 }{\sum_{j=1}^n \sigma_j^2 \, T_{j} \, |\textbf{A}^{+}_{j}|^2} |\hat{H}_+|^2 +  \frac{|\textbf{A}^{\times}|^4 }{\sum_{k=1}^n \sigma_k^2 \, T_{k} \, |\textbf{A}^{\times}_{k}|^2} |\hat{H}_\times|^2 \,.
\end{equation}
where $\sigma_j^2$ and $T_j$ are the variance of the data distribution in a frequency band around $f_{gw}$ (usually few tenths of Hz wide) and the observation time in the j-th detector. The matched filter $\hat{H}_+$ is defined as (the same for $\hat{H}_\times$):
\begin{equation}\label{estim5n}
\hat{H}_+ =\frac{\textbf{X} \cdot \textbf{A}^{+}}{|\textbf{A}^{+}|^2}=\frac{\sum\limits_{j=1}^{n}\textbf{X}_j \cdot (\textbf{A}_j^{+})^*}{\sum\limits_{k=1}^{n}\textbf{A}_k^{+}\cdot(\textbf{A}_k^{+})^*}= \frac{1}{|\textbf{A}^{+}|^2} \left( |\textbf{A}_1^{+}|^2\cdot \hat{H}_{+,1}+...+|\textbf{A}_n^{+}|^2\cdot \hat{H}_{+,n} \right)
\end{equation}
In $\hat{H}_{+/\times}$, each data 5-vector “interacts" only with the corresponding template. 

In the hypothesis of Gaussian noise with zero mean and variance $\sigma^2_j$, the corresponding components of the data 5-vector are also distributed according to a complex Gaussian distribution with mean value zero and variance $\sigma_j^2\cdot T_{j}$. 
Therefore, $\hat{H}_{+/\times}$ have also Gaussian distributions,
\begin{equation}
\hat{H}_{+/\times}\sim Gauss\left(x;\, 0,\sigma^2_{+/\times}\right)
\qquad \text{with} 
\qquad
\sigma^2_{+/\times}=\sum\limits_{j=1}^{n}\frac{\sigma^2_j\cdot T_j \cdot |\textbf{A}_j^{+/\times}|^2}{|\textbf{A}^{+/\times}|^4}
\end{equation}Since $|\hat{H}_{+/\times}|^2=\textit{Re}[\hat{H}_{+/\times}]^2+\textit{Im}[\hat{H}_{+/\times}]^2$, it follows that $|\hat{H}_{+/\times}|^2\sim Exp(x; \sigma^2_{+/\times})$ and $S \sim \Gamma(x;2,1)$.

Let us consider the toy case of a network of $n$ co-located ($\,|\textbf{A}_k^{+/\times}|^2=|\textbf{A}^{+/\times}_1|^2 \,,\forall \, k$) detectors with the same observation time $t$. In this case, the variances $\sigma^2_{+/\times}$ are: 
\begin{equation}\label{aritmean}
    \sigma_{+/\times}^2=\frac{t}{|\textbf{A}^{+/\times}_1|^2} \frac{\sum_{j=1}^n\sigma^2_j}{n}
\end{equation}This essentially means that we are treating the entire network as if it were a single detector, with a variance equal to the arithmetic mean of the variances of the individual detectors. It's important to note that even in this scenario, the multidetector analysis does not necessarily outperform the most sensitive detector alone.

\section{Weighted multidetector extension}\label{best5n}
Since the noise levels and the observation times in the detectors can be different, the standard 5n-vector could reduce the signal-to-noise ratio compared to the 5-vector of the most sensitive detector. In this Section, we define the "weighted" data 5n-vector as:
\begin{equation}\label{5nvec_weight}
\mathbf X=[c_1\mathbf X_1,..., c_n\mathbf X_n] 
\end{equation} where the weights $c_j$ are:
\begin{equation}
c_j = \frac{\sqrt{n}}{\sqrt{\sum\limits_{i=1}^n\left( \frac{T_i}{\sigma^2_i}  \right)} }\sqrt{\frac{T_j}{\sigma^2_j}}= \sqrt{\mathcal{H}} \cdot \sqrt{\frac{T_j}{\sigma^2_j}}
\end{equation}and $\mathcal{H}$ is  the harmonic mean of the time-weighted variances. 

The matched filters are (the same for $\hat{H}_\times$):
\begin{equation}
\hat{H}_+ =\frac{\textbf{X} \cdot \textbf{A}^{+}}{|\textbf{A}^{+}|^2}=\frac{\sum\limits_{j=1}^{n}c_j \textbf{X}_j \cdot (\textbf{A}_j^{+})^*}{\sum\limits_{k=1}^{n}\textbf{A}_k^{+}\cdot(\textbf{A}_k^{+})^*}= \frac{1}{|\textbf{A}^{+}|^2} \left(c_1 |\textbf{A}_1^{+}|^2 \hat{H}_{+,1}+...+ c_n |\textbf{A}_n^{+}|^2 \hat{H}_{+,n} \right)
\end{equation}
In the hypothesis of Gaussian noise for the j-th detector with variance $\sigma_j^2$, the two complex estimators $\hat{H}_{+/\times}$ have also Gaussian distributions,
\begin{equation}\label{variance}
\hat{H}_{+/\times}\sim Gauss\left(x;\, 0,\sigma^2_{+/\times}\right)
\qquad\text{with} 
\qquad
\sigma^2_{+/\times}=\sum\limits_{j=1}^{n}\frac{(c_j\sigma_j)^2\cdot T_j \cdot |\textbf{A}_j^{+/\times}|^2}{|\textbf{A}^{+/\times}|^4}
\end{equation}
  Using the $c_j$, we re-define the noise variance in each detector:
\begin{equation}
(c_j\sigma_j)^2  = \frac{n \cdot T_j}{\sum\limits_{i=1}^n\left( \frac{T_i}{\sigma^2_i} \right) }
\end{equation}
 If the observation time was the same ($\forall \, j,\,T_j=t $), this corresponds to "equalize" the noise in each detector. In this case, it follows that:
 \begin{equation}
\sigma^2_{+/\times}=\sum\limits_{j=1}^{n}\frac{(c_j\sigma_j)^2\cdot T_j \cdot |\textbf{A}_j^{+/\times}|^2}{|\textbf{A}^{+/\times}|^4}=\frac{n \cdot t}{\sum\limits_{i=1}^{n} \left(\frac{1}{\sigma_i^2}\right)\sum\limits_{k=1}^{n} |\textbf{A}_k^{+/\times}|^2}
\end{equation}

Considering the toy case previously discussed, the variances are 
\begin{equation}
\sigma^2_{+/\times}=\frac{t}{\sum\limits_{i=1}^{n} \left(\frac{1}{\sigma_i^2}\right) |\textbf{A}^{+/\times}_1|^2}
\end{equation}This corresponds to consider one detector with observation time $t$ and variance $V^2$ given by:
\begin{equation}
V^2=\frac{1}{\sum\limits_{i=1}^{n} \left(\frac{1}{\sigma_i^2}\right)}=\frac{\mathcal{H}}{n}
\end{equation} 
Here, $\mathcal{H}$ represents the harmonic mean of the variances of the individual detectors. Since there is the condition:
\begin{equation}\label{imp}
\frac{min\{\sigma_1^2,...,\sigma^2_n\}}{n}\leq \frac{\mathcal{H}}{n} \leq  min\{\sigma_1^2,...,\sigma^2_n\} \,
\end{equation}
it follows that for $n$ co-located detectors with the same observation time, using the coefficients $c_j$ always results in an improvement in detection sensitivity. This is different from what was found in the classic definition of the 5n-vector, as shown in \eqref{aritmean}. 

It is evident that in the general case, with detectors located at different positions and having varying observation times, a multi-detector analysis may not necessarily outperform the most sensitive detector.

\section{Conclusion}
In conclusion, the standard multidetector extension of the 5-vector method does not consistently yield better sensitivity when compared to the most sensitive individual detector.

In this letter, we have introduced a novel multidetector extension of the 5-vector method, taking into account the detectors sensitivities and their respective observation times. Our analytical results, based on a simplified scenario involving $n$ co-located detectors, demonstrate that the weighted extension consistently outperforms the standard extension, as opposed to the case of using the most sensitive detector alone.

However, in the more general context of having $n$ different detectors, it remains unclear under which conditions a multidetector analysis outperforms the most sensitive single detector analysis. In the next future, a criterion that considers the source's sky position, detector sensitivities, and observation times should be analyzed. This criterion will help determine when and which detectors should be employed to effectively enhance sensitivity in CW targeted searches.

\bibliographystyle{unsrtnat}
\bibliography{biblio}

\end{document}